\documentclass[a4paper, 10pt, conference]{ieeeconf} 
\AtBeginDocument{\let\autocite\cite}

\IEEEoverridecommandlockouts                              

\usepackage{balance}
\usepackage{url}
\usepackage{cite}
\usepackage{graphicx} 
\usepackage{flushend}
\usepackage{booktabs}
\usepackage{tikz}
\usepackage{tabularx}

\newcommand\copyrighttext{%
  \footnotesize \textcopyright 2026 IEEE. 
  Permission from IEEE must be obtained for all uses, in any current or future
  media, including reprinting/republishing this material for advertising or promotional
  purposes, creating new collective works, for resale or redistribution to servers or
  lists, or reuse of any copyrighted component of this work in other works.}
\newcommand\copyrightnotice{%
\begin{tikzpicture}[remember picture,overlay]
\node[anchor=south,yshift=10pt] at (current page.south) {\fbox{\parbox{\dimexpr\textwidth-\fboxsep-\fboxrule\relax}{\copyrighttext}}};
\end{tikzpicture}%
}

\begin{document}
\IEEEoverridecommandlockouts
\overrideIEEEmargins

\title{\LARGE \bf
  Designing Task-Induced Arousal: A Multimodal Stress Induction Method for Interactive Experiments
}

\author{Morten Roed Frederiksen$^{1}$
  \thanks{{$^{1}$Morten Roed Frederiksen {\tt\small mrof@itu.dk} is affiliated with the Data Systems \& Robotics department at the IT-University of Copenhagen Denmark.}}
}

\maketitle
\copyrightnotice
\begin{abstract}
HCI and HRI studies often require short, repeatable arousal manipulations that can run while participants continue interacting with a device or robot. These experiments are often challenged by the need to induce arousal in settings that still resemble real interaction. Participants must continue using a device, touching a robot, or producing sensor data while the manipulation unfolds. We aimed to develop a compact and repeatable way to induce controlled task-related arousal during interactive experiments by combining a lightweight browser-based pacing task, escalating timing demands and urgency cues, and a concurrent physical hotwire-style challenge. In an A-B-A within-participant study, the induction condition significantly increased mental demand, temporal demand, effort, frustration, and SAM arousal (all $p < .001$), while perceived performance decreased ($p < .001$). GSR peak rate increased relative to both calm conditions ($p = .030$) and escalated over time ($p < .001$). Grip variability also increased ($p = .036$), as did release speed (both $p < .001$), while valence remained above the scale midpoint. These results provide initial evidence that the combined procedure induces controlled, relatively high-valence task-related arousal and may serve as a reusable experimental tool for future human-computer and human-robot interaction studies.
\end{abstract}
\section{Introduction}

Controlled stress induction is central to psychophysiology, affective computing, and human-performance research because it allows researchers to study how cognitive, physiological, and behavioral systems respond to increased demands under laboratory conditions \cite{Kirschbaum1993TheS,Healey2005DetectingSD,Picard1997AffectiveC}. Widely used paradigms include social-evaluative stressors such as the Trier Social Stress Test, physical stressors such as the cold pressor test, and exposure-based paradigms involving public speaking or phobia-specific stimuli \cite{Kirschbaum1993TheS,Hines1936TheCP,Pertaub2002AnEO,Miloff2019AutomatedVR}. These methods are powerful, but they are not always suitable for short HCI and HRI studies: they may be time-consuming, ethically demanding, dependent on social threat, physically aversive, or tied to specific fears \cite{Frisch2015TheTS,Miloff2019AutomatedVR}.
HCI and HRI studies often require stressors that are short, repeatable, easy to integrate with sensors, and compatible with ongoing interaction \cite{Healey2005DetectingSD,Norden2022InducingAR}. The participant may need to continue touching a device, holding a robot, or producing behavioral data while the arousal manipulation unfolds, constraints that some established paradigms were not designed to meet. The distinction between threat-based stress and challenge-based arousal is also important. Challenge stressors can raise arousal and effort without producing strongly negative affect, making them more suitable for interactive studies that need to remain ethically straightforward and repeatable \cite{Horan2020ARO,Podsakoff2023LayingTF}.
\begin{figure}[t]
    \centering
    \includegraphics[width=0.9\linewidth]{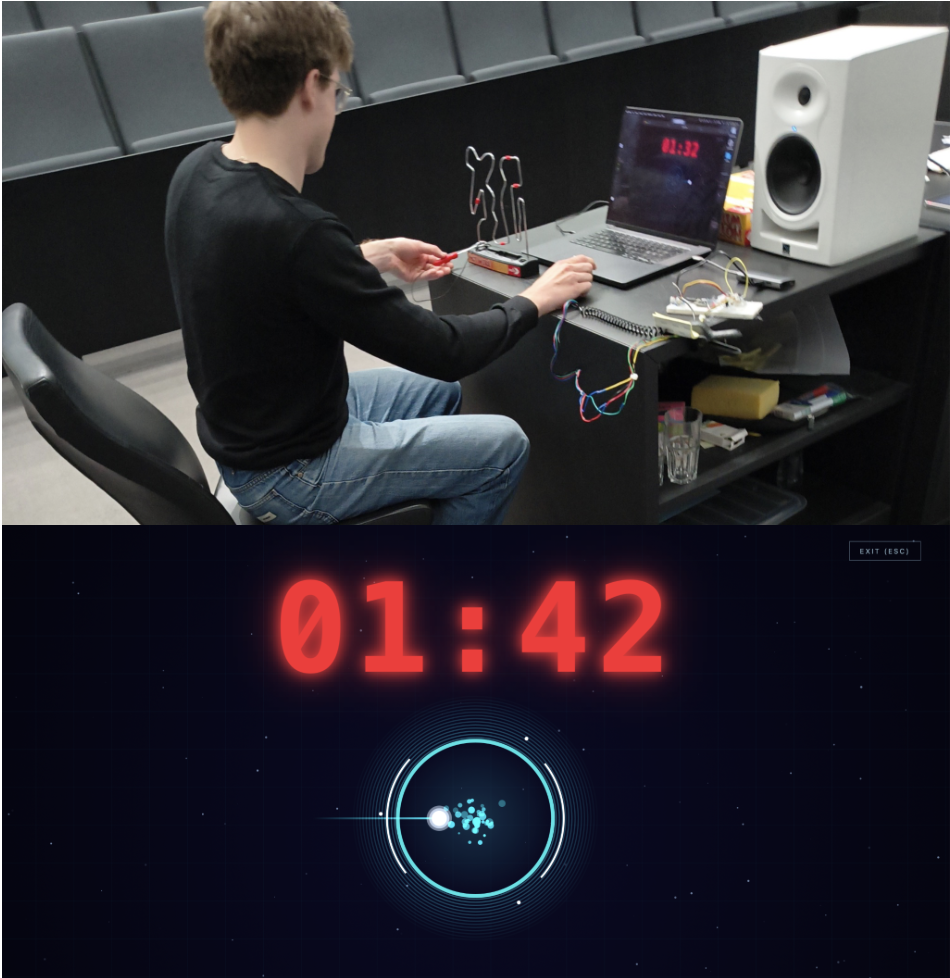}
    \caption{
        \textbf{Experimental setup and screen content during the arousal-induction condition.}
        \textit{Top:} A participant performing the B condition. The right hand holds the handheld device and presses in response to the moving target on the laptop screen, while the left hand navigates the physical hotwire track. A loudspeaker to the right delivers auditory urgency cues as the countdown progresses.
        \textit{Bottom:} The browser-based task display during the same condition. A white point orbits the central ring at increasing speed; the participant must press the handheld device each time the point passes through the target zone.
    }
    \label{fig:setup}
\end{figure}
This paper presents a compact, open-source arousal induction method designed for short laboratory studies of interactive systems \cite{AffectaStressSoftware2026}. The method combines a standardized browser-based timing task with a concurrent physical hotwire style challenge. The browser component is implemented as a lightweight single-file web application that controls the timing, pacing, visual cues, and auditory escalation, whereas the hotwire apparatus supplies an embodied divided-attention task. The software runs directly in a standard browser alongside different sensing systems and requires no specialized experimental software. Reproducibility therefore depends on both the shared browser implementation and adequate specification of the accompanying physical task. \cite{Horan2020ARO,Podsakoff2023LayingTF}.
The method was evaluated in an A-B-A within-participant study using subjective self-report, electrodermal activity, and grip-pressure dynamics \cite{Bradley1994MeasuringET,Hart1988DevelopmentON}. The induction condition produced large increases in mental demand, temporal demand, effort, and SAM arousal (all of these $p<.001$), while GSR peak rate escalated progressively across the two-minute window in a pattern that tracked the task's escalating structure. Valence decreased modestly but remained above the scale midpoint, consistent with challenge-based rather than distress-based activation. Together, these results suggest that the procedure is a reusable building block for HCI and HRI studies requiring controlled arousal during ongoing interaction. The browser component provides a common pacing and cue structure across platforms, while replication of the complete multimodal procedure additionally requires a physically matched concurrent task.
\section{Related Work}

Stress induction is well established in psychophysiology, but many validated paradigms were not designed for interactive-system experiments. The Trier Social Stress Test induces robust psychobiological stress through anticipatory pressure, public speaking, and mental arithmetic before an evaluative audience \cite{Kirschbaum1993TheS}, and the cold pressor test induces acute physiological stress through sustained pain from cold-water immersion \cite{Hines1936TheCP}. Both are powerful but require either an evaluative panel or a physical apparatus that is incompatible with concurrent handheld interaction. Shorter task-based alternatives have been proposed: the Sing-a-Song Stress Test uses anticipated vocal performance to elicit autonomic responses within a compact protocol \cite{Brouwer2014ANP}, and the Digital Stress Test adapts arithmetic and verbal-performance demands to a smartphone format for scalable, multimodal stress induction outside the laboratory \cite{Norden2022InducingAR}. More recent work has explored speech improvisation under increasing cognitive and social demands as a further alternative \cite{Saskovets2025ValidationOA}. In HRI, psychophysiological measures have been proposed as a way to evaluate user responses to robot behavior, presence, and interaction context \cite{Tiberio2013PsychophysiologicalMT,Bethel2007SurveyOP}. Recent HRI datasets further show the value of combining physiological signals with affect labels in robot interaction scenarios, but also highlight the need for clear timing, labeling, and synchronization between interaction events and physiological data \cite{Heinisch2024PhysiologicalDF}. These requirements make stress induction a methodological problem: the manipulation must be strong enough to alter arousal, but not so disruptive that it prevents the interaction or contaminates the sensing context.
A related issue is how the induced state is characterized. Workload instruments such as NASA-TLX are widely used because task difficulty, time pressure, effort, and perceived performance are central to human-performance evaluation \cite{Hart1988DevelopmentON,Hart2006NasaTaskLI}. Affective instruments such as the Self-Assessment Manikin separately measure valence and arousal, allowing researchers to distinguish activation from negative affect \cite{Bradley1994MeasuringET}. This distinction is important because challenge-based stressors, including time pressure, divided attention, and escalating task difficulty, can increase workload and arousal without producing strongly negative valence \cite{Horan2020ARO,Podsakoff2023LayingTF}. Dual-task and divided-attention paradigms have long been used to increase cognitive load and task-related stress in controlled settings \cite{Galy2012WhatIT}, and the combination of a primary timing task with a concurrent fine-motor task creates exactly this kind of demand profile: high workload and urgency without social evaluation or pain. For interactive-system research, this implies that a useful induction method should make clear whether it induces workload-related challenge, low-valence distress, or social-evaluative threat, a distinction that self-report batteries combining NASA-TLX and SAM can jointly address.
The present work addresses this methodological gap by introducing a portable, challenge-based induction procedure for interactive experiments. Unlike paradigms that require an evaluative panel, a separate smartphone protocol, or a task disconnected from the interaction object, the proposed method is designed to run while participants continue interacting with a handheld device. It combines escalating timing demands, visible urgency, auditory cues, divided attention, and a tangible performance challenge, while preserving compatibility with continuous physiological and behavioral sensing. The contribution is methodological rather than a new robotic system or stress-detection model. It provides an induction procedure that can be incorporated into HCI and HRI experiments in which participants must continue touching, controlling, or otherwise interacting with a device or robot while arousal is manipulated. Its relevance to HRI therefore lies in experimental control and measurement design during physical interaction with a robotic device, rather than in the introduction of a new autonomous robot capability.

\section{Methods}

\subsection{Study Design}

The study used a within-participant A--B--A design to evaluate whether a compact task-based procedure could induce short-term task-related arousal in a laboratory interaction setting. The first condition, A1, served as a calm baseline. The second condition, B, introduced the arousal-induction procedure. The third condition, A2, repeated the calm task in order to examine whether responses returned toward baseline after the induction phase.

The primary methodological question was whether the B condition produced a reliable increase in task-related arousal relative to the calm interaction periods. The validation strategy was deliberately multimodal. The induction was evaluated using subjective workload and affect ratings, physiological electrodermal activity, and behavioral grip-pressure dynamics. This was done to determine whether the method produced converging evidence of increased workload and arousal rather than relying on a single measurement channel.

\subsection{Participants}

Twenty-six participants took part in the experiment. Four participants were not included in the complete-case analyses because one or more required recordings were missing or incomplete. The final complete-case dataset therefore contained 22 participants with usable A1, B, and A2 recordings. The sample was intended to provide an initial within-participant manipulation check rather than a definitive estimate of population-level effects or physiological variability. Participants were recruited through voluntary participation at the university and included both children and university students. Ages ranged from 10 to 44 years. Participants received a soda for participating.

For child participants, parental consent was obtained, and the children provided oral assent before participation. The induction procedure was designed to be difficult, time-pressured, and frustrating, but also game-like. It therefore avoided pain, phobia-specific content, formal social evaluation, and deception-based threat.

\subsection{Apparatus}

Participants interacted with a pocket-sized handheld tactile device, hereafter referred to as the handheld device. The device contained two side-button pressure sensors and integrated electrodes for galvanic skin response measurement. Participants held the device in their dominant hand throughout the experimental tasks and pressed the side buttons in response to visual timing cues. The same device therefore functioned as the interaction object and as a sensor platform.

Sensor data were logged during each condition. Each row contained timestamps, pressure readings from the two side-button sensors, and GSR readings. The firmware attempted to record one row every 50 ms, corresponding to an intended sampling rate of approximately 20 Hz. The GSR signal used in the analysis was the short-window averaged value produced on the device during recording.

The visual timing task was presented on a separate screen using a browser-based program. The browser task provided the pacing, visual cues, countdown, and auditory escalation used during the experimental conditions. Participant responses were recorded through the handheld device rather than through the browser interface.

\subsection{Browser Task and Open-Source Design}

The stress/arousal induction program was implemented as a single-file web application using HTML, CSS, JavaScript, and the Canvas API \cite{AffectaStressSoftware2026}. The program displayed a full-screen visual environment with a central target circle and a small glowing point moving horizontally across the screen. Participants were instructed to squeeze or press the handheld device when the moving point entered the central target.

The program contained two selectable modes. The calm mode lasted 60 s and was used for the A1 and A2 conditions. The stress/arousal mode lasted 120 s and was used for the B condition. Both modes used the same basic visual timing task, which ensured that the required hand action remained comparable across conditions. The main difference was that the B condition added time pressure, increasing pace, sensory escalation, and a concurrent physical task.

In the calm mode, the point moved at a constant speed across the screen. The central target used a calm cyan visual style, and no red countdown timer was shown. In the stress mode, the same point movement was retained, but the speed increased progressively over the two-minute condition. The movement speed ramped linearly from the baseline speed to approximately 2.6 times the baseline speed by the end of the condition. This increased the required response pace as the condition progressed.

The stress mode also displayed a large red countdown timer starting at 02:00. The timer remained visible throughout the B condition and updated continuously. Auditory cues were triggered when 60 s, 30 s, and 10 s remained. At 30 s remaining, the countdown display began flashing, and the final part of the condition included additional alarm-like urgency. The central target also changed visually during the stress condition, shifting from the cyan baseline display toward a redder color profile as time progressed. Thus, the browser task combined temporal escalation, visual urgency, and auditory salience within a controlled two-minute procedure.

The browser task did not function as a performance-scoring system. Instead, it acted as a standardized pacing and induction environment. This design allowed the task to be paired with the handheld tactile device, which recorded grip pressure and GSR throughout the interaction. The separation between the browser-based pacing environment and the sensing device allows the procedure to be paired with different HCI and HRI setups, although replication of the full induction also requires a comparable concurrent physical task.

The induction software was designed to be lightweight, inspectable, and reusable. Because it runs in a standard web browser, it does not require specialized experimental software, a specific operating system, or a particular robot platform. The implementation explicitly defines the condition durations, speed ramp, countdown behavior, cue timing, visual state changes, and session transitions. These parameters can therefore be inspected, replicated, or modified by other researchers.

This is important for HCI and HRI because interactive-system studies often differ in their sensing hardware, robotic platforms, task context, and physical setup. The browser component can provide a standardized timing and cue sequence while the device or robot under study remains responsible for interaction and measurement. The same induction procedure can therefore be paired with handheld robots, wearable sensors, desktop interfaces, mobile devices, or other interaction platforms.

Open sourcing the system also improves methodological transparency. Rather than describing the stressor only as ``increasing time pressure'' or ``escalating sound,'' the exact implementation can be shared with the research community. Researchers can reproduce the original timing profile or adapt specific components, such as duration, speed ramp, auditory cues, or visual intensity, to match their own ethical constraints and study design. This makes the browser implementation a transparent and modifiable component of the tested induction procedure and a building block for future interactive-system studies.

\subsection{Experimental Procedure and Conditions}

Participants were introduced to the study as an investigation of biofeedback and tactile interaction using a small handheld robot. They first completed a pre-study questionnaire based on Self-Assessment Manikin (SAM) estimating pre-study valence/arousal. Then they were allowed to hold and view the handheld device and the onscreen visual timing task. Participants were instructed to hold the device in their dominant hand and to press or squeeze the side buttons when the moving point entered the central target circle on the screen.

All three experimental conditions used the same basic timing task. In A1 and A2, participants completed the calm one-minute version of the task. In B, participants completed the two-minute arousal-induction version while also performing a concurrent hotwire-style task with their non-dominant hand. Small pauses occurred between conditions while participants completed self-report items.

In the calm A1 and A2 conditions, participants performed the browser-based timing task for 60 s. The point moved at a constant pace, the display remained in the calm visual state, relaxing background audio was played, and the red countdown timer was not shown. A1 provided the initial calm baseline. A2 repeated the same task after the induction to examine whether subjective, physiological, and behavioral measures returned toward the calm-task level.

In contrast, the B condition lasted 120 s and was designed to induce task-related arousal through escalating workload, time pressure, divided attention, and repeated performance disruption. Participants continued to perform the same visual timing task with the handheld device, but the browser program increased the pace of the moving point over time. The speed ramp required participants to respond more frequently as the condition progressed.

The B condition also introduced explicit urgency cues. A large red countdown timer was shown throughout the condition. Auditory cues were triggered at key remaining-time thresholds, and the final part of the condition included a flashing countdown and additional alarm-like sound. The visual display also became more urgent over time, with the target color shifting toward red during the stress mode. These elements were designed to make the time limit salient and to increase perceived urgency without introducing social evaluation or physical discomfort.

During the same condition, participants completed a difficult hotwire-style task with their non-dominant hand. Participants guided a conductive handheld loop along the track. The track contained bends or directional changes, with a minimum loop-to-track clearance of approximately 6 mm on each side. The base measured 15 × 7 cm, and the apparatus was positioned 50 cm from the participant. Contact closed an electrical circuit and activated a distracting buzzer tone. Participants began at the left-side starting position, moved toward the right-side end position, and returned to the starting position after each contact. The apparatus and participant position were held constant across sessions.

This required participants to divide attention between the screen-based timing task, the handheld tactile device, and the physical hotwire task. If the wire was touched, a loud buzz was triggered, and participants were instructed to restart the hotwire task. None of the participants completed the hotwire task successfully during the B condition.


\subsection{Validation Measures}

Subjective validation was used to examine whether the B condition increased perceived workload and activation relative to the calm baseline condition. Workload was assessed using raw NASA-TLX dimensions. The included dimensions were mental demand, temporal demand, perceived performance, effort, and frustration. Mental demand assessed perceived cognitive load. Temporal demand assessed time pressure. Performance assessed perceived success in completing the task; lower scores therefore indicate poorer perceived performance. Effort assessed how hard participants had to work. Frustration assessed feelings of insecurity, discouragement, irritation, stress, or annoyance.

Affective state was assessed using the Self-Assessment Manikin. SAM arousal was used to examine whether participants felt more activated during the B condition. SAM valence was used to examine whether the induction produced a strongly negative affective state or whether the experience remained closer to challenge-related activation. The expected pattern was an increase in workload and arousal, together with a possible reduction in valence that would not necessarily indicate a strongly aversive or threat-based state.

For each subjective measure, participant-level condition scores were used in the within-participant analyses. The main subjective validation question was whether the B condition differed from the calm baseline condition in the expected direction.

Physiological validation was based on electrodermal activity as a measure of arousal. Because the electrodes were integrated into a device that participants actively held and pressed, the analysis did not treat the absolute GSR level as a clean tonic measure. Absolute level could be affected by grip pressure, finger placement, movement, and contact quality. The analysis therefore focused on phasic GSR peak structure within participants.

For each participant and condition, phasic GSR peaks were detected from the averaged GSR signal. Peak detection used an adaptive prominence threshold scaled to the standard deviation of the GSR signal within each participant-condition segment. Peak rate was calculated as the number of detected peaks divided by condition duration in minutes. The main physiological validation question was whether peak rate increased during B relative to A1 and A2. The B condition was also examined over time to determine whether GSR peak rate increased as the induction escalated. Electrodermal activity was the only autonomic measure available in the device used in the experiment. The physiological analysis therefore evaluates sympathetic electrodermal activation specifically and does not provide a broader cardiovascular characterization of stress.

Behavioral validation was based on grip-pressure dynamics. The two side-button pressure readings were combined into a single grip-pressure signal for each participant and condition. The analysis focused on dynamic features of the grip signal rather than on mean pressure alone, because the expected stress-related effect was not a simple increase in maximum grip strength.

Grip variability was calculated within each participant and condition as the variability of the combined pressure signal over time. Release dynamics were estimated from decreases in the pressure signal, with faster downward changes interpreted as faster release behavior. These metrics were used to test whether the B condition altered the temporal structure of gripping and releasing during interaction.

The behavioral validation question was whether the arousal-induction condition produced more variable grip behavior and faster release dynamics than the calm baseline conditions.
\begin{figure}[t]
    \centering
    \includegraphics[width=\linewidth]{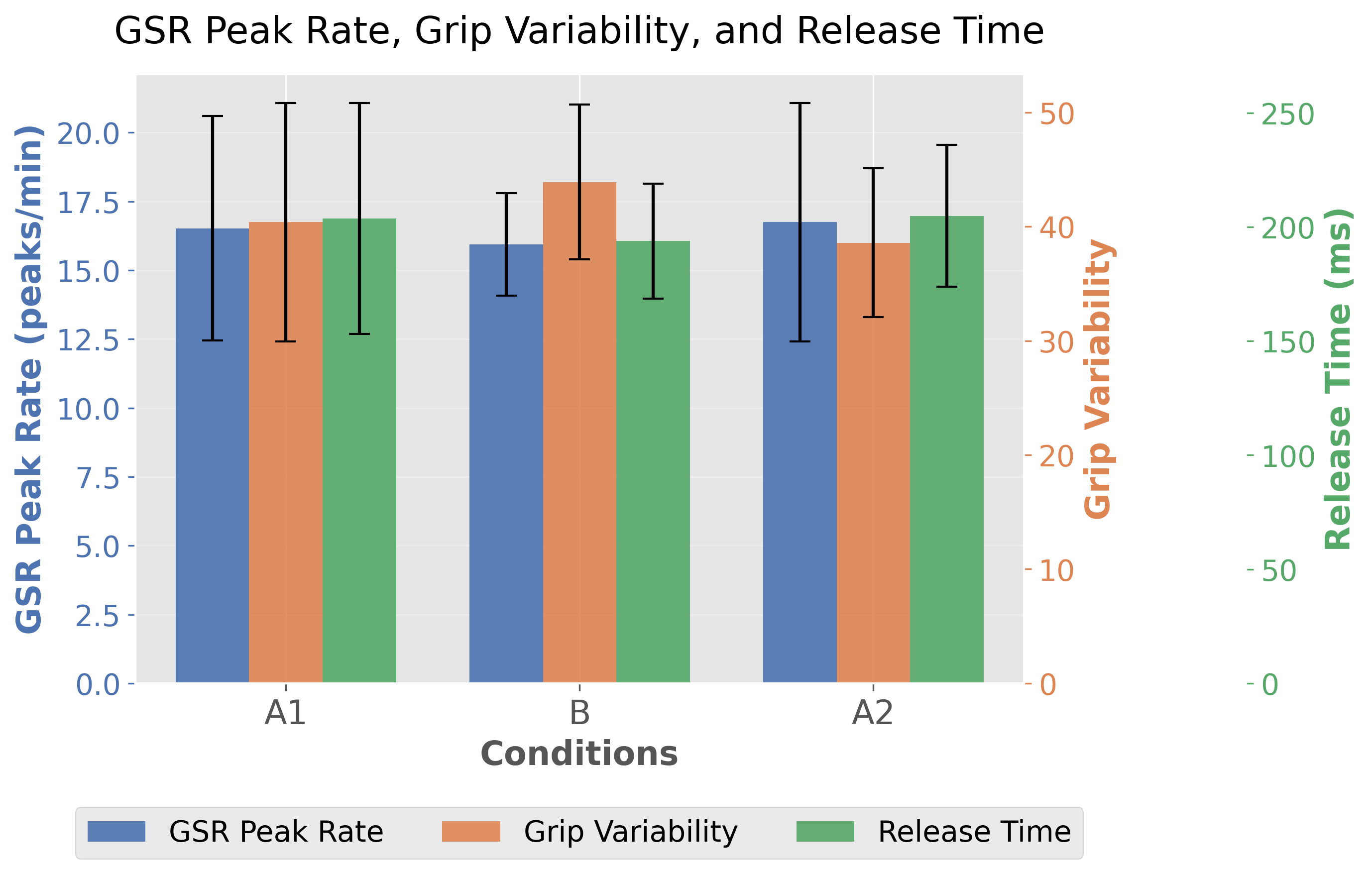}
    \caption{Objective validation measures across experimental conditions. Bars show GSR peak rate, grip variability, and release speed across A1, B, and A2. The B condition showed increased phasic GSR activity, higher grip variability, and faster release dynamics relative to the calm conditions. Error bars show standard deviation.}
    \label{fig:objective_validation}
\end{figure}
\subsection{Validation Criterion}

The method was evaluated as a multimodal induction procedure rather than as a single-measure manipulation. The B condition was considered successful if it produced converging evidence across three domains: increased subjective workload and arousal, increased phasic electrodermal activity, and altered grip-pressure dynamics. This criterion was chosen because the method was designed to induce short-term task-related arousal in an interactive setting, not to produce a purely fear-based, painful, or socially evaluative stress response.


\subsection{Statistical Analysis}

All statistical analyses used within-participant comparisons. A1--B comparisons were used to test whether the induction increased subjective workload, subjective arousal, physiological arousal, and grip-dynamic change relative to the initial calm condition. Where relevant, B--A2 comparisons were used to examine whether the same measures were also higher during the induction than during the post-induction calm condition.

Paired-samples t-tests were used for condition comparisons. Effect sizes were reported as Cohen's $d_z$, calculated from the paired differences. Statistical significance was evaluated at $\alpha = .05$.

\section{Results}

All analyses were conducted on the complete-case dataset of 22 participants with usable A1, B, and A2 recordings. The results are organized according to the three validation domains of the method: subjective self-report, physiological electrodermal activity, and behavioral grip-pressure dynamics. The central question was whether the B condition produced a convergent increase in task-related arousal relative to the calm baseline conditions.

\begin{figure}[t]
    \centering
    \includegraphics[width=\linewidth]{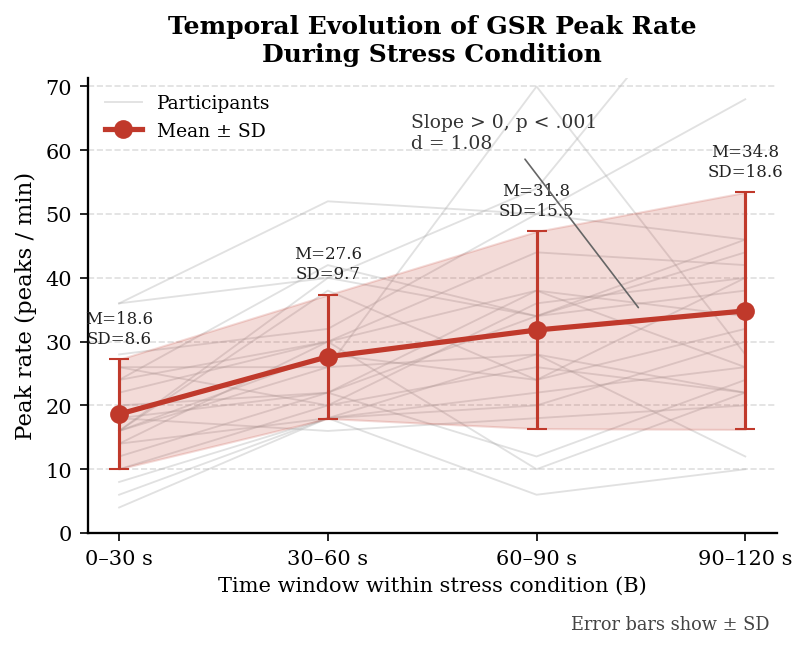}
    \caption{
    \textbf{Temporal evolution of GSR peak rate during the stress/arousal condition.}
    The red line shows the mean GSR peak rate across participants in four 30 s windows of the stress-inducing condition. Grey lines show individual participant trajectories across the same windows. 
    }
    \label{fig:gsr_temporal_evolution}
\end{figure}

\subsection{Subjective Validation}

The subjective measures showed a clear and consistent effect of the arousal-induction condition. As summarized in Table~\ref{tab:subjective_results}, participants experienced the B condition as substantially more demanding than the calm baseline condition. Mental demand, temporal demand, effort, and frustration all increased significantly from A1 to B. Perceived performance decreased significantly, indicating that participants experienced the B condition as more difficult to perform successfully.

The largest effects were observed for temporal demand, effort, mental demand, and perceived performance. This pattern matches the intended design of the induction procedure: the B condition was constructed to increase time pressure, cognitive demand, divided attention, and repeated performance disruption. Frustration also increased significantly, but the effect was smaller than the effects for time pressure, effort, and mental demand. Thus, the subjective workload results indicate that the manipulation primarily increased task demand and effort rather than only producing irritation or negative affect.

\begin{table}[t]
\centering
\caption{Subjective validation of the arousal-induction condition. The table reports the primary A1--B manipulation check.}
\label{tab:subjective_results}
\resizebox{\columnwidth}{!}{%
\begin{tabular}{lcccc}
\hline
Measure & A1 $M$ ($SD$) & B $M$ ($SD$) & Main change & $p$ \\
\hline
Mental demand & 1.82 (1.01) & 5.95 (1.13) & Increased & $<.001$ \\
Temporal demand & 1.45 (0.86) & 5.64 (1.05) & Increased & $<.001$ \\
Effort & 2.18 (1.05) & 6.27 (0.94) & Increased & $<.001$ \\
Frustration & 1.77 (1.41) & 4.50 (1.77) & Increased & $<.001$ \\
Performance & 6.14 (0.89) & 2.45 (1.34) & Decreased & $<.001$ \\
SAM arousal & 4.32 (1.89) & 6.68 (1.84) & Increased & $<.001$ \\
SAM valence & 6.82 (1.68) & 5.45 (1.92) & Decreased & .005 \\
\hline
\end{tabular}%
}
\vspace{0.5em}
\begin{minipage}{0.95\columnwidth}
\vspace{0.5em}
\footnotesize
Note. All subjective ratings were scored on 1--7 scales. Higher NASA-TLX scores indicate more of the named dimension. For performance, higher scores indicate better perceived performance; therefore, the decrease from A1 to B indicates poorer perceived task success during the induction condition.
\end{minipage}
\end{table}

SAM ratings supported the same interpretation. Arousal increased significantly during B, showing that participants felt more activated during the induction condition. Valence decreased significantly from A1 to B, but the mean valence rating during B remained above the midpoint of the scale. After the induction, arousal decreased again and valence increased again in A2. This pattern suggests that the induction produced a temporary increase in activation and workload, accompanied by a moderate reduction in pleasantness, rather than a strongly negative affective state.

\begin{figure*}[t]
    \centering
    \includegraphics[width=\textwidth]{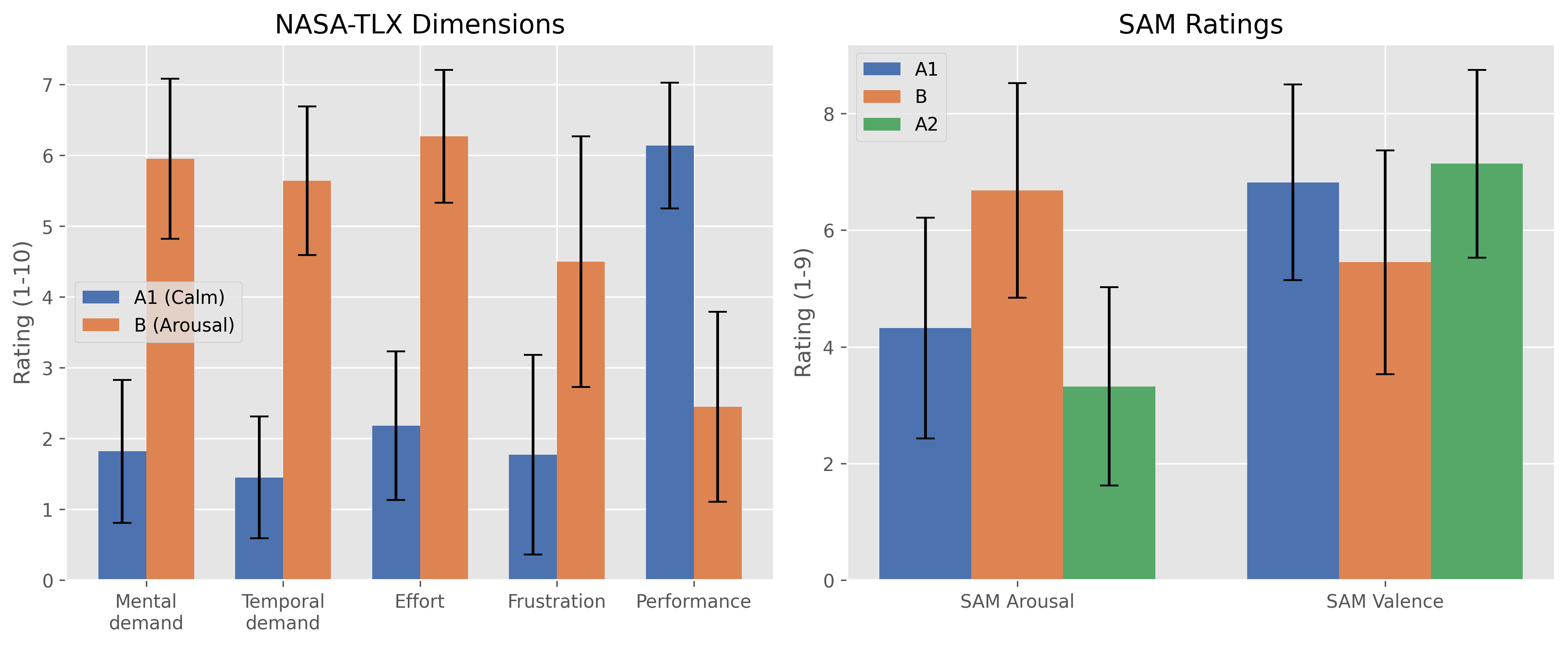}
    \caption{Subjective validation of the arousal-induction condition. NASA-TLX dimensions showed large increases in mental demand, temporal demand, effort, and frustration, while perceived performance decreased. SAM arousal increased during B, whereas SAM valence decreased but remained above the scale midpoint.}
    \label{fig:subjective_validation}
\end{figure*}

\subsection{Physiological Validation}

The physiological validation focused on phasic GSR activity because the electrodes were embedded in a handheld device that participants actively held and pressed. The main result was that the induction was reflected in the frequency of phasic GSR responses, not in the amplitude of individual peaks.

Relative GSR peak amplitude did not differ significantly across conditions. Mean peak amplitude was similar in A1, B, and A2, and none of the paired comparisons were significant. This indicates that the B condition did not produce larger individual GSR peaks.

In contrast, GSR peak rate increased during the B condition. As shown in Fig.~\ref{fig:objective_validation}, peak rate was significantly higher in B than in both A1 and A2, while the two calm conditions did not differ. This indicates that the stress/arousal condition produced more frequent phasic electrodermal responses rather than larger single responses.

The temporal structure of the GSR response also followed the design of the induction. Peak rate increased across the four 30-s windows of the B condition, and participant-level slopes were significantly greater than zero. The second minute of B also showed higher peak rate than the first minute. Thus, the physiological response did not merely appear at the onset of the B condition; it increased as the task became faster, more urgent, and more difficult.

\subsection{Behavioral Grip Validation}

Grip-pressure dynamics provided a behavioral validation of the induction. Mean grip pressure did not differ significantly across conditions, showing that the B condition did not simply make participants press harder overall. This is important because it suggests that the behavioral effect was not a simple increase in average force.

Instead, the induction affected the temporal dynamics of gripping and releasing. Grip variability increased during B relative to both calm conditions, and release speed was substantially higher during B than during A1 or A2. The two calm conditions did not differ significantly for these measures. This pattern indicates that the B condition changed how participants interacted with the device: their grip became less stable, and their release movements became faster.

Pairwise comparisons confirmed the pattern shown in Fig.~\ref{fig:objective_validation}. GSR peak rate was higher in B than in A1 and A2, while A1 and A2 did not differ. Grip variability was higher in B than in both calm conditions, and release speed showed the same pattern with a larger effect. Mean grip pressure did not differ between A1 and B. Thus, the behavioral signature of the induction was not stronger pressing, but more variable and faster grip-release dynamics.

\subsection{Convergent Validation of the Induction Method}

Across measures, the B condition produced a coherent validation pattern. Subjective ratings showed that participants experienced the task as more mentally demanding, time pressured, effortful, frustrating, and difficult to perform. SAM ratings showed increased arousal and a moderate reduction in valence, while valence remained above the midpoint of the scale. In the recorded electrodermal channel, GSR peak rate increased during B and escalated over time. Behaviorally, grip variability and release speed increased, whereas mean grip pressure did not. These grip measures reflect behavior on the present handheld device and should not be assumed to transfer directly to interaction systems with different physical forms or sensing arrangements.


\subsection{Exploratory Cross-Measure Associations}

As a secondary analysis, we examined whether participants who experienced the B condition as more demanding also showed stronger physiological or behavioral responses. These analyses were exploratory and were not used as primary validation tests. To avoid overinterpreting post-hoc associations, we report only the two correlations most directly connected to the design logic of the induction procedure.

First, the subjective increase in mental demand was associated with the temporal escalation of the physiological response. Participants who reported a larger increase in mental demand from A1 to B also showed a steeper increase in GSR peak rate across the four 30 s windows of the B condition, $\rho = .54$, $p = .009$. This supports the interpretation that the escalating structure of the B condition was not only present in the task design, but was also reflected in both subjective workload and physiological arousal.

Second, subjective workload was associated with behavioral grip dynamics. For this exploratory analysis, we calculated a workload-change composite by averaging the A1-to-B changes in mental demand, temporal demand, effort, frustration, and reverse-scored perceived performance. Performance was reverse-scored so that higher values consistently represented greater experienced task burden. The composite was used only to reduce multiple closely related post-hoc correlations and to represent the common directional workload response targeted by the induction; it was not treated as a validated NASA-TLX total score. The composite was positively correlated with release speed during B, $\rho = .50$, $p = .017$. Participants who experienced the induction as more demanding therefore tended to show faster release dynamics while interacting with the handheld device. Because the dimensions were combined for an exploratory convergence analysis rather than for confirmatory scale construction, this association should be interpreted cautiously and warrants replication in a larger sample. 

Together, these exploratory associations support the convergent validation of the induction method. The first association links the escalating task design to physiological arousal over time, while the second links perceived workload to behavioral change during handheld interaction. Because these analyses were exploratory, they should be interpreted as supportive rather than as independent evidence for the effectiveness of the induction.

\section{Discussion}
The present study provides an initial evaluation of a procedure for inducing task-related arousal during interaction with a handheld device. The method should not be interpreted as inducing stress in the broadest sense. Rather, it provides a controlled way to increase workload, urgency, and physiological activation while preserving compatibility with interactive-system studies, making it relevant for HCI and HRI contexts where participants must engage with a device or robot while an arousal manipulation unfolds. The present evaluation involved a small handheld tactile robot rather than an autonomous or socially interactive robotic platform. The contribution is therefore not a new robot capability, but a methodological procedure for manipulating arousal while participants continue physically interacting with a robotic device and producing continuous sensor data. Evaluation with other robotic embodiments and interaction contexts will be needed to establish how broadly the procedure transfers within HRI.

\subsection{Characterizing the Induced State}
The response pattern is most consistent with challenge-based arousal. Participants reported higher mental demand, temporal demand, effort, frustration, and arousal during the B condition, and these subjective changes were accompanied by physiological and behavioral shifts. Although valence decreased, the mean rating remained above the scale midpoint: the task became less pleasant, but not strongly negative. The paradigm should therefore not be described as inducing fear, anxiety, or low-valence distress; it induces a high-arousal challenge state characterized by workload, urgency, and performance pressure.
This distinction defines both the value and the boundary of the method. A high-valence challenge stressor is ethically easier to justify, repeatable, and compatible with playful robots, handheld devices, and game-like interaction. However, it should not be used to validate systems that claim to detect anxiety, fear, or distress, since a model responding to this paradigm may be detecting workload-related arousal rather than negative affect. Future work should examine whether variants can induce lower-valence stress while preserving the method's brevity and compatibility with interactive tasks.
\subsection{Scope and Limitations}
The physical hotwire task introduces a methodological trade-off. It increases setup complexity and may reduce standardization across laboratories, since physical builds can vary in sensitivity and feedback intensity. However, it also provides a concrete, goal-directed challenge that creates embodied performance pressure distinct from abstract cognitive tasks or artificial feedback. The task appears achievable but repeatedly produces interruption, which may help explain why the procedure raised workload and arousal without relying on pain, social evaluation, or deception.
The physical implementation should therefore be understood as both a strength and a limitation. Future development will integrate the hotwire component into the browser environment as an on-screen version, preserving the goal-directed failure structure while improving standardization, logging, and portability.

The grip-based results should be interpreted as device- and sample-specific behavioral indicators rather than as direct measures of stress. Grip variability and release dynamics may be influenced by hand size, finger placement, dominant-hand characteristics, baseline grip strength, age, motor control, and the fit of the handheld device. These characteristics were not systematically measured or normalized in the present study. 

The sample and physiological measures further constrain the scope of the conclusions. The complete-case sample of 22 participants limits the precision and generalizability of the findings, particularly for electrodermal measures that show substantial inter-individual variability. The broad age range and inclusion of both children and university students may introduce additional heterogeneity that the present sample was not large enough to model. Heart rate and heart-rate variability were not recorded. Their inclusion would provide a broader assessment of autonomic response and help determine whether the induction affects cardiovascular regulation in addition to electrodermal activity. Future evaluations should combine electrodermal measures with HR or HRV and, where feasible, respiration or other autonomic signals.

\section{Conclusion}

This paper presented a compact, browser-based method for inducing task-related arousal during interactive experiments. The method combines an escalating visual timing task with auditory urgency cues and a concurrent hotwire-style challenge, while preserving compatibility with handheld sensing and interaction. We conducted an initial A-B-A within-participant evaluation and found that the arousal-induction condition increased perceived workload and arousal, increased phasic GSR peak rate, and altered grip dynamics through higher variability and faster release behavior. These findings provide initial evidence that the tested procedure can induce controlled, relatively high-valence task-related arousal during handheld interaction. The lightweight browser component is reusable, while replication of the complete procedure additionally requires a comparable physical divided-attention task. Larger studies involving other robotic embodiments, autonomic measures such as HR or HRV, and more diverse interaction devices are needed before the procedure can be considered broadly validated across HRI contexts.

\balance

\bibliography{bibliography}
\bibliographystyle{IEEEtran}

\end{document}